\documentclass[reprint, a4paper, amsfonts, amssymb, aps, prl, footinbib
]{revtex4-1}
\usepackage[utf8]{inputenc}

\usepackage{amsmath}
\usepackage{graphicx,bm}
\usepackage[T1]{fontenc}
\usepackage{ae}
\usepackage{hyperref}

\graphicspath{{graphics/}}

\newcommand{\absq}[1]{\left\vert #1 \right\vert^2}
\newcommand{\abs}[1]{\left\vert #1 \right\vert}

\newcommand{\sgn}[1]{\mathrm{sgn}\left( #1 \right)}

\begin{document}
\title{Magnetic resonance in a singlet-triplet Josephson junction}
\author{Lars Elster}
\author{Manuel Houzet}
\author{Julia S. Meyer}
\affiliation{Univ.~Grenoble Alpes, INAC-SPSMS, F-38000 Grenoble, France \\ CEA, INAC-SPSMS, F-38000 Grenoble, France}
\email[]{lars.elster@cea.fr}

\date{\today}

\begin{abstract}
We study a singlet-triplet Josephson junction between a conventional s-wave superconductor and an unconventional p$_{\rm x}$-wave superconductor. The Andreev spectrum of the junction yields a spontaneous magnetization in equilibrium. This allows manipulating the occupation of the Andreev levels using an ac Zeeman field. The induced Rabi oscillations manifest themselves as a resonance in the current-phase relation. For a {circularly} polarized magnetic field, we find a spin selection rule, yielding Rabi oscillations only in a certain interval of the superconducting phase difference. 
\end{abstract}

\maketitle


The current-phase relation of a Josephson junction contains information about the Andreev levels and their occupations. Junctions formed between unconventional superconductors have exotic bound states leading to unusual current-phase relations. Amongst unconventional Josephson junctions, those realized between singlet and triplet superconductors are of special interest, because of their incompatible spin pairing symmetries. Their equilibrium properties have been studied for various types of heterogeneous junctions \cite{kwon_2004b, asano_2001, yoshida_1999, asano_2011, burset_2014, lu_2009}. 

Let us consider such a Josephson junction between a conventional spin-singlet, s-wave superconductor and an unconventional spin-triplet, p$_x$-wave superconductor. 
This junction displays exotic spin properties. Namely, it hosts two spin-polarized Andreev bound states, which have the same spin \cite{sengupta_2008}. In equilibrium, this results in a spontaneous magnetization 
that is $2\pi$-periodic in the superconducting phase difference. On the other hand, 
a $\pi$-periodic equilibrium supercurrent, which does not probe the exotic spin properties, is found \cite{yang_2011, yip_1993, asano_2003}. The spin properties of the Andreev levels open the possibility for spin manipulation, using a time-dependent Zeeman field. A similar idea, the manipulation of the Andreev levels in spin active Josephson junctions between conventional superconductors, has already been reported \cite{michelsen_2008}.

In this article, we show, that an ac Zeeman field leads to coherent Rabi oscillations between different spin states of the singlet-triplet junction. These Rabi oscillations manifest themselves as resonances in the current-phase relation. For a {circularly} polarized magnetic field, we find a spin selection rule, yielding Rabi oscillations only in a certain interval of the superconducting phase difference. The applied Zeeman field also induces non-coherent transitions between the bound states and the continuum states. In principle, these transitions, which we treat within a macroscopic master equation approach, could give rise to a decay mechanism for the Rabi oscillations. However, we find, that due to spin and energy constraints, these processes do not coexist with the Rabi oscillations.

\begin{figure}[tb]
\centering
\includegraphics[width=0.6\linewidth]{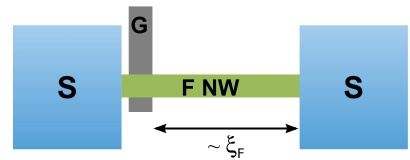}
\caption{Setup of an effective singlet/triplet junction using a semiconducting ferromagnetic nanowire (F NW) contacted with conventional singlet superconductors (S). The gate (G) allows for realizing a barrier with tunable transparency. By adjusting the length of the nanowire between the gate and the right superconductor, one can realize an effective triplet superconducting reservoir.}
\label{Fig:setup}
\end{figure}

A possible experimental realization of our proposal could be based on the quasi-onedimensional (TMTSF)$_2$X Bechgaard salts \cite{jerome_1980} in which a p$_x$-pairing was discussed, as suggested in Ref.~\cite{sengupta_2008}. Alternatively, we propose to realize a junction between conventional superconductors separated by a ferromagnetic semiconducting nanowire, as illustrated in Fig.~\ref{Fig:setup}. The gate would allow for the realization of a barrier with tunable transparency. Furthermore, an effective p$_x$-wave superconductivity is realized when the length of the nanowire between the gate and one of the leads matches the coherence length $\xi_{\rm F}$ for the superconducting correlations induced in the nanowire.

Let us introduce the model. The Hamiltonian describing a Josephson junction between an s-wave superconductor and a one-dimensional, time-reversal symmetric p$_{\rm x}$-wave superconductor reads
\begin{equation}
\label{eq:H0}
H = \int dx \; \Psi^{\dagger} {\cal H}\Psi\ ,
\end{equation}
where $\Psi = (R_{\downarrow}, L^{\dagger}_{\uparrow}, L_{\downarrow}, R^{\dagger}_{\uparrow})^T$, and $R^{\dagger}_{\sigma}$ and $L^{\dagger}_{\sigma}$ are creation operators for right-moving  and left-moving electrons with spin $\sigma=\uparrow,\downarrow$, respectively. The Bogoliubov-de~Gennes Hamiltonian ${\cal H}$ is given as 
\begin{equation}
\label{eq:HBdG}
{\cal H} = v_{\text{F}} p \eta_z\tau_z+  U(x) \eta_x\tau_z-\Delta_{\rm s}(x)\tau_x
+ \Delta_{\rm p}(x)  \eta_z \tau_xe^{-i \tau_z \phi} \ ,
\end{equation}
where $\tau_{x,y,z}$ and $\eta_{x,y,z}$ denote Pauli matrices in particle-hole and R/L spaces, respectively.
The first term in Eq.~\eqref{eq:HBdG}, with Fermi velocity $v_{\rm F}$ and momentum operator $p$, is the kinetic energy. The second term describes normal backscattering induced by a scalar potential $U(x)$ that vanishes outside the central region of the junction, $0<x<L$, where $L$ is the junction length. It gives rise to an electronic transmission probability $T$, when the junction is in the normal state. The third term describes s-wave pairing with gap $\Delta_{\rm s}(x)=\Delta_{\rm s}\theta(-x)$, where $\theta$ is the Heavside step function, on the left side of the junction. The last term describes p$_{\rm x}$-wave pairing with gap $\Delta_{\rm p}(x)=\Delta_{\rm p}\theta(x-L)$ between electrons having opposite spins along $z$-direction on the right side of the junction. Without loss of generality, $\Delta_{\rm s}$ and $\Delta_{\rm p}$ are assumed to be real. For simplicity, we will restrict our analysis to the case $\Delta_{\rm s}=\Delta_{\rm p}\equiv \Delta$ \footnote{When $\Delta_{\rm s}\neq\Delta_{\rm p}$, our conclusions remain valid in the regime of superconducting phase difference where two Andreev bound states exist, cf.~Ref.~\cite{sengupta_2008}.}.
The superconducting phase difference across the junction is denoted $\phi$. Note that we use units where $\hbar = 1$.

\begin{figure}[tb]
\centering
\includegraphics[width=0.9\linewidth]{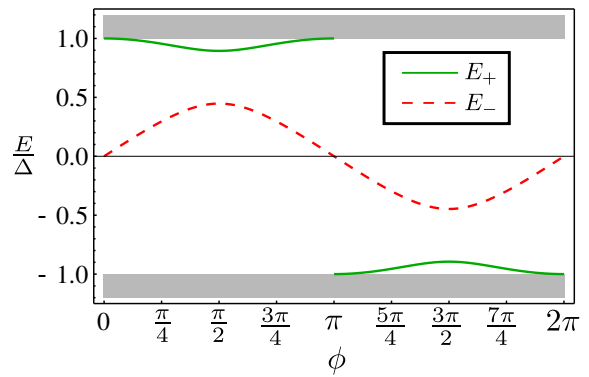}
\caption{Energy-phase relation of the two bound states for the transmission $T=0.8$. The continuum of states is indicated in gray. All states shown have spin $\downarrow$.}
\label{fig:ABS}
\end{figure}

In the short-junction limit, $L\ll v_{\rm F}/\Delta$, we find that the Bogoliubov-de~Gennes Hamiltonian \eqref{eq:HBdG} admits for two bound states with energies~\cite{sengupta_2008}
\begin{equation}
\label{ABS-energy}
E_\pm=\frac{\mathrm{sgn}(\sin\phi)}{\sqrt{2}}\Delta\sqrt{1 \pm \sqrt{1-T^2\sin^2\phi}}\ ,
\end{equation}
and wavefunctions $\psi_\pm(x)$ that are given in the Supplemental Material~\footnote{See Supplemental Material for details on the derivation of the eigenstates of Hamiltonian \eqref{eq:HBdG}, the stationnary solution of the master equation \eqref{eq:masterequation}, and the ionization and refill rates \eqref{eq:rates}.}. Note that the choice of the spinor $\Psi$ implies that we are considering states with spin down only. Furthermore, Eq.~\eqref{eq:HBdG} admits for a four-fold degenerate continuum of (outgoing) propagating states with energies $E$ ($|E|>\Delta$) and wavefunctions $\psi_{E\mu}(x)$, where $\mu$ is a degeneracy index. Using a Bogoliubov transformation, 
\begin{equation}
\Psi(x) = \sum_{\nu=\pm} \psi_{\nu}(x) \gamma_{\nu} + \sum_{E,\mu} \psi_{E\mu}(x) \gamma_{E\mu}\ ,
\label{eq:bdg_trafo}
\end{equation}
where $\gamma_{\nu}$ and $\gamma_{E\mu}$ are annihilation operators for quasiparticles in the bound state with energy $E_{\nu}$ and for quasiparticles in the continuum with energy $E$ and degeneracy index $\mu$, respectively, we may diagonalize the Hamiltonian \eqref{eq:H0} to obtain
\begin{equation}
\label{eq:Hdiag}
H= \sum_{\nu=\pm} E_{\nu} \gamma^{\dagger}_{\nu} \gamma_{\nu} +\sum_{E,\mu} E \gamma^{\dagger}_{E\mu} \gamma_{E\mu}\ .
\end{equation}
A typical spectrum is shown in Fig.~\ref{fig:ABS}. Note that at vanishing coupling, $T\to 0$, the spectrum of the s-wave lead is gapped, while the p$_{\rm x}$-wave lead, which realizes two copies of the Kitaev model \cite{kitaev_2001} in opposite spin sectors, admits for a zero-energy edge state. A finite coupling moves this state to finite energy and yields the bound state $\nu=-$, while a second bound state ($\nu=+$) detaches from the continuum. In contrast to conventional junctions, both bound states carry the same spin ($\sigma=\downarrow$).

The bound state occupations, $n_{\nu}=\langle \gamma^{\dagger}_{\nu} \gamma_{\nu} \rangle$, determine the magnetization carried by the junction, 
\begin{equation}
M=-\frac{\mu_{\text{B}}}{2}\sum_{\nu=\pm}  \left(n_{\nu}-\frac 12\right)\ ,
\end{equation}
where $\mu_{\text{B}}$ is the Bohr magneton. In equilibrium, $n_\nu=f(E_\nu)$, where $f$ is the Fermi function.  As a result, the junction carries a spontaneous magnetization, which is $2\pi$-periodic in the phase difference~\cite{sengupta_2008}. 

The Josephson current is given as
\begin{equation}
I = 2e\sum_{\nu=\pm} \frac{d E_{\nu}}{d \phi} \left(n_{\nu}-\frac 12\right)\ .
\label{eq:current}
\end{equation}
The equilibrium supercurrent is spin-insensitive and, therefore,  it is $\pi$-periodic.

Thus, to probe the peculiar spin properties of the junction, we have to 
consider out-of-equilibrium effects. In particular, in order to manipulate the bound state occupations, we apply an ac magnetic field, which is described by the Zeeman Hamiltonian
\begin{equation}
\label{eq:HZ}
 H_{\text{Z}} = \mu_{\text{B}} \sum_{s,s'=\uparrow,\downarrow} \int dx\; \bm{B}\cdot \left(R^\dagger_{s}\bm{\sigma}_{ss'} R_{s'}+L^\dagger_{s} \bm{\sigma}_{ss'} L_{s'}\right)\ .
\end{equation}
We assume the magnetic field to be uniform and circularly polarized in the plane perpendicular to the spin quantization axis, $\bm{B}=B(\cos \Omega t, \sin \Omega t, 0)$, where $\Omega$ is the driving frequency. Such a field leads to spin-flip processes. In particular,  the spin of the system changes by  $\Delta S_z=\sgn{\Omega}$ when a photon is absorbed, whereas it changes by $\Delta S_z=-\sgn{\Omega}$ when a photon is emitted.  In the following, we concentrate on the case $\Omega<0$.

In order to identify the processes induced by the field, we express Eq.~\eqref{eq:HZ} in terms of the quasiparticle operators using the Bogoliubov transformation \eqref{eq:bdg_trafo}. We find
\begin{eqnarray}
H_{\text{Z}} &=& \mu_{\text{B}} Be^{-i \Omega t} 
\left(V_{+,-} \gamma_+ \gamma_- + \sum_{E;\mu,\nu} V_{\nu,E\mu} \gamma_{\nu} \gamma_{E\mu}  
\right.
\nonumber\\
&&\left.
+ \frac{1}{2} \sum_{E,E';\mu,\mu'} V_{E\mu,E'\mu'}\gamma_{E\mu}\gamma_{E'\mu'}\right) + \mathrm{h.c.}\ ,
\label{eq:HBqp}
\end{eqnarray}
where $V_{\lambda,\lambda'}=\int dx \;\psi_{\lambda}^T \eta_x (-i\tau_y)\psi_{\lambda'}$ for $\lambda\in\{+,-,E\mu\}$. 

According to Eq.~\eqref{eq:HBqp}, the field couples two quasiparticle states. Three different types of processes are possible: transitions involving only bound states [first term of Eq.~\eqref{eq:HBqp}], transitions involving a bound state and a continuum state [second term of Eq.~\eqref{eq:HBqp}], and transitions involving only continuum states [third term of Eq.~\eqref{eq:HBqp}]. 

For the discussion of the spin properties, we note that the destruction of a quasiparticle with spin down at negative energies corresponds to the creation of a quasiparticle with spin up at positive energies. 
So far, we used both positive and negative energies for spin $\downarrow$. In the following, we will work with both spin directions, but only positive quasiparticle energies.
Furthermore, we will assume that the temperature is low (on the scale of $\Delta$), such that the continuum states at $E>\Delta$ are empty. 

\begin{figure}[tb]
\centering
\includegraphics[width=\linewidth]{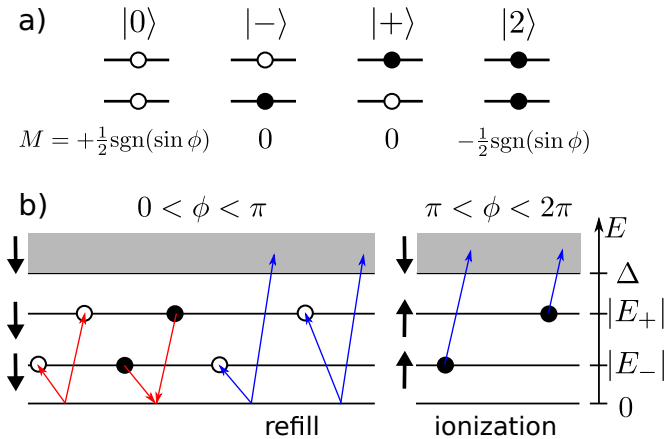}
\caption{a) The four possible states of the junction and their magnetization $M$. Full (open) dots represent occupied (empty) states. b) Transitions induced by a circularly polarized magnetic field with $\Omega < 0$. The shaded region is the continuum of states. The thick black arrows denote the spin of the states. Absorption (emission) of a photon changes the spin by $\Delta S_z = -1$ ($\Delta S_z = +1$). The Andreev bound states carry a spin down (up) in the phase interval $0<\phi<\pi$ ($\pi<\phi<2\pi$). }
\label{fig:processes}
\end{figure}

The transitions involving only bound states correspond to Rabi oscillations, i.e., coherent oscillations between the state $|0\rangle$, where both bound states are empty, and the state $|2\rangle$, where both bound states are occupied. Rabi oscillations occur when the oscillation frequency $|\Omega|$ matches the Rabi frequency, $\Omega_{\rm R}=|E_+(\phi)+E_-(\phi)|$. Using the energy dispersions given by Eq.~\eqref{ABS-energy}, the interesting frequency regime is, thus, given by $\Delta<|\Omega|<\sqrt2\Delta$~\footnote{Note that as $T$ decreases the maximal value of $|E_+(\phi)+E_-(\phi)|$ decreases.}. Then, sweeping the phase at fixed frequency, the resonance condition is met for 4 different values of the phase: $\phi_0,\pi-\phi_0,\pi+\phi_0,2\pi-\phi_0$. However, the spin selection rule imposes a further constraint. Namely, at $\Omega<0$, Rabi oscillations are possible only if the bound states carry a spin down, which is the case in the interval $0<\phi<\pi$. Thus, the circularly polarized magnetic field leads to Rabi oscillations only at two values of the phase: $\phi_0$ and $\pi-\phi_0$.

Transitions involving a bound state and a continuum state change the parity of the bound state occupation, namely they connect the even-parity subspace $\{|0\rangle,|2\rangle\}$ to the odd-parity subspace $\{|-\rangle,|+\rangle\}$, where $|\nu\rangle$ denotes the state in which only the bound state with energy $|E_\nu|$ is occupied. (For a sketch of the four states of the junction, see Fig.~\ref{fig:processes}a).) This would represent a decay mechanism for the Rabi oscillations. We may distinguish two different processes, that are sketched in Fig.~\ref{fig:processes}b).

In an {\em ionization} process, a particle from a bound state is promoted to a continuum state. Energy conservation imposes the condition $|\Omega| > \Delta-|E_\nu|$ for such a process. In the frequency range of interest for Rabi oscillations, this condition is always met. However, the spin selection rule imposes that, at $\Omega<0$, the bound state carries a spin up, which is the case in the interval $\pi<\phi<2\pi$ only. Thus, Rabi oscillations and ionization processes occur in different phase intervals.

In a {\em refill} process, a Cooper pair is broken such that one particle occupies a bound state, whereas the second particle is promoted to a continuum state. Here energy conservation imposes the condition $|\Omega| > \Delta+|E_\nu|$. In the frequency range of interest for Rabi oscillations, this condition is never met for the state with energy $|E_+|$. By contrast, for the state with energy $|E_-|$, one obtains a critical phase $\phi_{\rm{c}}$ such that the condition is met in the phase intervals $[-\phi_{\rm{c}},\phi_{\rm{c}}]$ and $[\pi-\phi_{\rm{c}},\pi+\phi_{\rm{c}}]$. Comparing $\phi_{\rm{c}}$ with the phase $\phi_0$ giving the resonance condition for Rabi oscillations, we find $\phi_{\rm{c}}<\phi_0$. Thus, Rabi oscillations and refill processes also occur in different phase intervals. Note that, here, the spin selection rule imposes that, at $\Omega<0$, the bound state carries a spin down, which is the case in the interval $0<\phi<\pi$.

We conclude that transitions between a bound state and the continuum due to a circularly polarized magnetic field do not provide a decay mechanism for the Rabi oscillations due to energy and spin constraints. However, such a decay may be due to other parity non-conserving processes related to, e.g., quantum phase fluctuations due to the resistive environment of the junction~\cite{olivares_2014, riwar_2014}.

Finally, transitions involving only continuum states have a threshold $|\Omega|>2\Delta$. Thus, they do not play a role in the frequency range of interest for Rabi oscillations.

The modifications of the bound state occupations, which are induced by the different processes discussed above, may lead to strong deviations of the Josephson current \eqref{eq:current} from its equilibrium value. To compute the Josephson current in the steady state, we introduce the matrix elements $\rho_{\alpha \beta} = \langle \alpha | \rho | \beta \rangle$ of the reduced density matrix $\rho$, where $|\alpha\rangle,|\beta\rangle\in\{|0\rangle,
|+\rangle,|-\rangle,|2\rangle\}$, so that Eq.~\eqref{eq:current} reads
\begin{equation}
I = (I_+ + I_-)(\rho_{00} - \rho_{22}) + (I_+ - I_-)(\rho_{--} - \rho_{++})\ ,
\label{eq:noneqcurrent}
\end{equation}
where $I_\pm=-e(d|E_\pm|/d\phi)$.

Taking into account the above considerations for the frequency range where Rabi oscillations can take place, we find the following behavior in different phase intervals. For phases $\phi\in[\pi,2\pi]$, only ionization processes are possible. Thus, the bound states are always empty, i.e.,  $\rho_{00}=1$ and $\rho_{--}=\rho_{++}=\rho_{22}=0$. As a consequence, the current remains equal to its equilibrium value, $I^{\rm eq}=I_++I_-$. On the other hand, in the phase intervals $[0,\phi_{\rm{c}}]$ and $[\pi-\phi_{\rm{c}},\pi]$, the ac field yields a refill process for the state $\nu=-$. Assuming that the rates for parity non-conserving processes due to the environment are much smaller than the field-induced rates, this state then will always be filled, i.e., $\rho_{--}=1$ and $\rho_{00}=\rho_{++}=\rho_{22}=0$. Thus, the current is $I=I_+-I_-$. 

To evaluate the current in the phase interval $[\phi_c,\pi-\phi_c]$, which includes the phases $\phi_0$ and $\pi-\phi_0$ where Rabi oscillations take place, we use the stationary solution of the master equation,
\begin{widetext}
\begin{equation}
\frac{d}{dt}\left(\begin{matrix} \rho_{00} \\ \rho_{22} \\\bar{\rho}_{02} \\ \bar{\rho}_{20}\\ \rho_{--} \\ \rho_{++}   \end{matrix}\right) = 
\left(\begin{matrix}
-\Gamma_-^{\rm R}-\Gamma_+^{\rm R} & 0 & i \frac{\omega_1^*}{2} & -i \frac{\omega_1}{2} &  \Gamma_{-}^{\text{I}} & \Gamma_{+}^{\text{I}}  \\
0 & -\Gamma_-^{\rm I}-\Gamma_+^{\rm I} & -i \frac{\omega_1^*}{2} & i \frac{\omega_1}{2} & \Gamma_{+}^{\text{R}} & \Gamma_{-}^{\text{R}} \\
i \frac{\omega_1}{2} & -i \frac{\omega_1}{2} & i \delta \omega -\frac{\Gamma_{\Sigma}}{2} & 0 & 0 & 0 \\
-i \frac{\omega_1^*}{2} & i \frac{\omega_1^*}{2} & 0 & -i \delta \omega -\frac{\Gamma_{\Sigma}}{2} & 0 & 0 \\
\Gamma_-^{\text{R}} & \Gamma_+^{\text{I}} & 0 & 0 & -\Gamma_-^{\text{I}} -\Gamma_+^{\text{R}} & 0\\
\Gamma_+^{\text{R}} & \Gamma_-^{\text{I}} & 0 & 0 & 0 & -\Gamma_-^{\text{R}} -\Gamma_+^{\text{I}} \\
\end{matrix}\right)
\left(\begin{matrix} \rho_{00} \\ \rho_{22} \\\bar{\rho}_{02} \\ \bar{\rho}_{20}\\ \rho_{--} \\ \rho_{++}   \end{matrix}\right).
 \label{eq:masterequation}
\end{equation}
\end{widetext}
Here, $\omega_1 = 2 V_{+,-} \mu_{\rm{B}} B$ with $\absq{V_{+,-}} =T^2\abs{\sin\phi}(1+\abs{\sin\phi})/(1+T \abs{\sin\phi})^2$, $\delta \omega = \Omega + \mathrm{sgn}(\sin\phi)\Omega_{\rm R}$, $\bar{\rho}_{02}=e^{i \Omega t} \rho_{02}$, and $\bar{\rho}_{20}=e^{-i \Omega t} \rho_{20}$. Furthermore, $\Gamma^{\rm I/R}_\nu$ are the ionization ($\rm I$) and refill ($\rm R$) rates of the state $\nu$, respectively, and $\Gamma_\Sigma=\sum_{\nu=\pm}(\Gamma_\nu^{\rm I}+\Gamma_\nu^{\rm R})$. The other 10 elements of the $4 \times 4$ density matrix that are not shown remain zero along the time-evolution. 

Solving Eq.~\eqref{eq:masterequation} to obtain the steady-state occupations~\cite{Note2}, we find that the current,
\begin{equation}
I = I^{\infty} + \frac{\Gamma^2}{\Gamma^2 + (2\delta\omega)^2} (I^0-I^{\infty}),
\label{eq-Iinf}
\end{equation}
is the sum of a background term $I^{\infty}$ and a resonant term.
The background term is given as $I^{\infty} = \sum_{\nu=\pm} I_{\nu}({\Gamma_{\nu}^{\rm I} - \Gamma_{\nu}^{\rm R}})/{\Gamma_{\nu}}$, where $\Gamma_{\pm} = \Gamma_{\pm}^{\rm I} + \Gamma_{\pm}^{\rm R}$. The current at resonance is given as
\begin{equation}
I^0 = \frac{\Gamma_\Sigma}{\Gamma^2} \left[\Gamma_\Sigma I^{\infty} 
+\frac{\absq{\omega_1}}{\Gamma_+\Gamma_-} (I_+ - I_-)  \sum_{\nu=\pm}\nu(\Gamma_\nu^{\rm I} - \Gamma_\nu^{\rm R})\right]\ ,
\label{eq-I0}
\end{equation}
whereas the width of the resonance is determined by 
\begin{equation}
\Gamma = \Gamma_{\Sigma}\sqrt{1+ \frac{\absq{\omega_1}}{\Gamma_+ \Gamma_-}}\ .
\label{eq-gamma}
\end{equation}
As pointed out above, all field-induced decay rates are zero in the phase interval $[\phi_{\rm{c}},\pi-\phi_{\rm{c}}]$. Therefore, we introduce phenomenological rates $\gamma$ to describe the parity non-conserving processes due to the environment. At low temperature, the refill processes are negligible \footnote{Such a process would require either an excess quasi-particle above the gap or a spin-flip process.}, and we are left with two rates $\gamma_\nu^{\rm I}$. In that case, $I^{\infty}$ reduces to the equilibrium current, $I^{\infty}=I^{\rm eq}$. Assuming $\gamma_\pm^{\rm I}\ll \abs{\omega_1}$, the current at resonance is obtained as
\begin{eqnarray}
I^0 
\approx \frac{\gamma_+^{\rm I} -\gamma_-^{\rm I}}{\gamma_+^{\rm I}+\gamma_-^{\rm I}} (I_+ - I_-)
\ ,
\end{eqnarray}
whereas the width of the resonance is
\begin{equation}
\Gamma  
\approx \frac{\gamma_+^{\rm I}+\gamma_-^{\rm I}}{\sqrt{\gamma_+^{\rm I} \gamma_-^{\rm I}}}\abs{\omega_1}.
\end{equation}
As the state $+$ is closer to the continuum, we expect $\gamma_+^{\rm{I}} \geq \gamma_-^{\rm{I}}$. Depending on their relative magnitude,  
the current may be completely suppressed at resonance, when $\gamma_-^{\rm I}=\gamma_+^{\rm I}$, or change its sign as compared to the equilibrium current, reaching a magnitude $I^0\approx I_+-I_-$, when $\gamma_-^{\rm I}\ll\gamma_+^{\rm I}$.

\begin{figure}[tb]
\centering
{\includegraphics[width=0.9\linewidth]{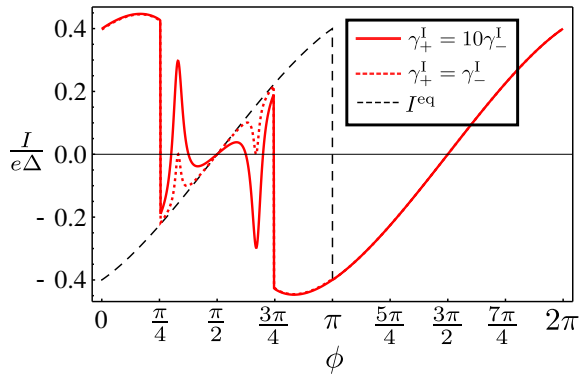}}
\caption{Current-phase relation for a junction with transmission $T=0.8$ and a circularly polarized magnetic field, with amplitude $\mu_{\rm B} B=10^{-2}\Delta$ and frequency $\Omega = -1.3 \Delta $. The phenomenological ionization rates are chosen as $\gamma_+^{\rm I}+\gamma_-^{\rm I}= 10^{-6} \Delta$ with $\gamma_{-}^{\rm I}=0.1\gamma_{+}^{\rm I}$ (solid red line) and $\gamma_{-}^{\rm I}=\gamma_{+}^{\rm I}$  (dotted red line). The equilibrium current is given for comparison (dashed black line).}
\label{fig:circular}
\end{figure}

Fig.~\ref{fig:circular} shows the non-equilibrium current-phase relation for a circularly polarized Zeeman field in the entire phase range. The $2\pi$-periodicity is due to the spin-sensitive manipulation of the bound state occupations. If the sign of $\Omega$ was reversed, the current-phase relation would be phase-shifted by $\pi$, i.e., $I(\Omega,\phi) = I(-\Omega,\phi + \pi)$.

By contrast, the spin sensitivity is lost, if we use a linearly polarized field, $\bm{B}=2B(\cos\Omega t, 0,0)$.
Such a field can be viewed as the superposition of two circularly polarized fields with opposite helicities.
Thus, there is no spin selection rule anymore, and
Rabi oscillations may now occur at 4 phases ($\phi_0,\pi-\phi_0,\pi+\phi_0,2\pi-\phi_0$). Furthermore, the field-induced ionization rates are non-zero for all superconducting phase differences while $\Gamma_-^{\rm R}$ exists in the phase intervals $[-\phi_{\rm{c}},\phi_{\rm{c}}]$ and $[\pi-\phi_{\rm{c}},\pi+\phi_{\rm{c}}]$. 

Extending the master equation \eqref{eq:masterequation} to the linear case and introducing a rotating-wave approximation to describe the vicinity of the Rabi resonances, we find that the steady-state current is given by Eqs.~(\ref{eq-Iinf})-(\ref{eq-gamma}) with $\Gamma_\nu^{\rm X}=\Gamma_\nu^{\rm X}(\Omega)+\Gamma_\nu^{\rm X}(-\Omega)$ , where ${\rm X}={\rm I}, {\rm R}$.

\begin{figure}[tb]
\centering
\includegraphics[width=0.9\linewidth]{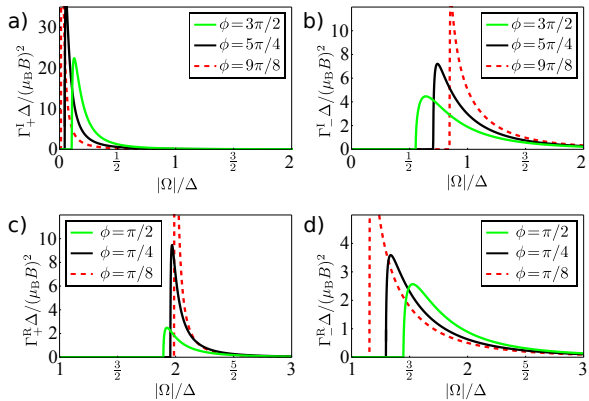}
\caption{Ionization (I) and refill (R) rates for $T=0.8$ as a function of the driving frequency $\Omega < 0$ for several phase differences $\phi$. a) Ionization rate of the state $\nu=+$. b) Ionization rate of the state $\nu=-$. c) Refill rate of the state $\nu=+$. d) Refill rate of the state $\nu=-$.}
\label{fig:rates}
\end{figure}

The transition rates for the ionization and refill processes involving the bound state $\nu$ can be calculated from Eq.~\eqref{eq:HBqp} using Fermi's Golden Rule,
\begin{eqnarray}
\! \Gamma^{{\rm I/R}}_{\nu} (\Omega)= 2\pi & (\mu_{\rm{B}} B)^2 \int_{\Delta}^{\infty} dE\; \rho(E) \sum_{\mu}  \absq{V_{\nu,\mp E\sgn{ \sin\phi}\mu}} \nonumber \\[1ex]
&\times \delta\left[\Omega + (|E_{\nu}| \mp E)\, \sgn{\sin\phi}\right].
\label{eq:rates}
\end{eqnarray}
Here $\rho(E)=(2\pi v_{\rm{F}})^{-1}{E}/{\sqrt{E^2-\Delta^2}}$ is the density of states in the leads.
Fig.~\ref{fig:rates} shows the frequency dependence of the ionization and refill rates. 
Note that the rates $\Gamma^{{\rm I/R}}_{\nu}$, whose typical amplitude is $\sim (\mu_{\rm B}B)^2/\Delta$, vanish below the threshold frequency $\Omega=\Delta\mp |E_\nu|$, as discussed above. Furthermore, they are suppressed at large frequencies $|\Omega|\gg\Delta$, while they display a maximum in the vicinity of the threshold frequency. Simplified expression for the rates in the ballistic ($T=1$) and opaque ($T=0$) limits are provided in~\cite{Note2}.

\begin{figure}[tb]
\centering
\includegraphics[width=0.9\linewidth]{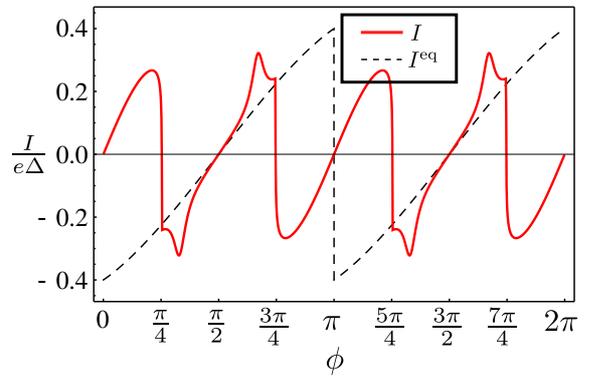}
\caption{Current-phase relation for a junction  with transmission $T=0.8$ and a linearly polarized magnetic field, with ${\mu_{\rm B} B}=10^{-2}\Delta$ and $\Omega = -1.3 \Delta $ (solid red line). The equilibrium current is given for comparison (dashed black line).}
\label{fig:linear}
\end{figure}

The current-phase relation for a linearly polarized field is shown in Fig.~\ref{fig:linear}. As the manipulation of the bound state occcupations is not spin sensitive in that case, the current is $\pi$-periodic as in equilibrium. Generically, the out-of-equilibrium current-phase relation is $2\pi$-periodic as soon as the ac field carries a finite angular momentum, leading to spin-dependent refill and ionization rates.

In conclusion, we have shown that the occupations of the Andreev levels in a Josephson junction between an s-wave and a  p$_{\rm x}$-wave superconductor can be manipulated using an ac Zeeman field. The induced Rabi oscillations manifest themselves as resonances in the current-phase relation. For a given circular polarization, their presence or absence depends on the spin state of the junction, thus providing a spin detection scheme.

\begin{acknowledgments}
LE thanks Roman-Pascal Riwar for helpful discussions. We acknowledge support by the AGIR program of the Universit\'e Grenoble-Alpes, by ANR through
grants ANR-11-JS04-003-01 and ANR-12-BS04-0016-03,
and by an EU-FP7 Marie Curie IRG.
\end{acknowledgments}

\bibliography{literature/references_final}

\end{document}


\begin{center}
\large \textbf{Supplemental Material: Magnetic resonance in a singlet/triplet Josephson junction}
\end{center}

The Supplemental Material provides technical details on some results that were used in the main text. In Sec.~\ref{sec:wavefunctions}, we derive the eigenstates of the Bogoliubov-de Gennes Hamiltonian, Eq.~(2) in the main text. In Sec.~\ref{sec:rates}, we use these eigenstates to compute the ionization and refill rates due to an ac Zeeman field, Eq.~(17) in the main text. In Sec.~\ref{sec:stationary}, we provide the stationary solution of the master equation, Eq.~(11) in the main text.

\section{wave functions}
\label{sec:wavefunctions}

In order to obtain the eigenstates of the Hamiltonian (2), we determine the general form of the wave functions in the leads in Sec.~\ref{sec:WFgeneral}. The wave functions are given in the basis $\eta \otimes \tau$, where $\eta$ denotes the R/L space and $\tau$ the particle-hole space. Then we use the boundary condition at the junction to establish the wave functions for the bound states in Sec.~\ref{sec:WFbound}, and for the continuum states in Sec.~\ref{sec:WFcontinuum}. We provide simple expressions both in the cases of a transparent and an opaque junction. As the wave functions are $2\pi$-periodic, we restrict our considerations to the interval $\phi \in [0,2\pi[$.

\subsection{Wave functions in the leads}
\label{sec:WFgeneral} 
In the left (s-wave) lead, $x<0$, the Hamiltonian (2) reduces to
\begin{equation}
{\cal H}_{\rm s} = v_{\text{F}} p \eta_z\tau_z -\Delta \tau_x.
\label{eq:sup:HS}
\end{equation}
It has a block-diagonal structure in the R/L space. In each block, characterized by $\eta_z=\pm1$, we thus need to solve an auxiliary $2\times2$ eigenvalue problem given by
\begin{equation}
(\pm v_{\text{F}} p \tau_z -\Delta \tau_x) \zweivector{u}{v} = E \zweivector{u}{v}.
\end{equation}
Using the solutions for this problem, we find that the most general form of the wave functions associated with the Hamiltonian \eqref{eq:sup:HS} at energies above the gap, $|E|>\Delta$, is the superposition of four independent spinors,
\begin{equation}
\psi(x) = \frac1{\sqrt{1+\alpha^2}}\left[A_{\rm e}^{\rm in} \viervector{1}{-\alpha}{0}{0}e^{i k x} + A_{\rm h}^{\rm out}\viervector{- \alpha}{1}{0}{0}e^{-i k x} + A_{\rm e}^{\rm out}  \viervector{0}{0}{1}{-\alpha}e^{-i k x}+A_{\rm h}^{\rm in} \viervector{0}{0}{-\alpha}{1}e^{i k x} \right].
\label{eq:sup:wavefunction_L}
\end{equation}
Here, $\alpha =(E-{\rm sign}(E)\sqrt{E^2-\Delta^2})/\Delta$ and $k={\rm sign}(E)\sqrt{E^2-\Delta^2}/v_{\rm F}$. The prefactor in Eq.~\eqref{eq:sup:wavefunction_L} ensures that each 4-spinor is normalized to unity. Furthermore, 
\begin{equation}
\frac 1{\sqrt{1+\alpha^2}}=\sqrt{\frac 12 \left(1+ \frac{\sqrt{E^2-\Delta^2}}{|E|}\right)}
\qquad {\rm and} \qquad
\frac \alpha{\sqrt{1+\alpha^2}}={\rm sign}(E)\sqrt{\frac 12 \left(1- \frac{\sqrt{E^2-\Delta^2}}{|E|}\right)}
\label{eq:sup:alpha}
\end{equation}
are nothing but the BCS coherence factors. Thus, the spinors with the coefficients $A_{\rm e}^{\rm in}$, $A_{\rm h}^{\rm out}$, $A_{\rm e}^{\rm out}$, and $A_{\rm h}^{\rm in}$ in Eq.~\eqref{eq:sup:wavefunction_L} describe right-moving electron-like, left-moving hole-like, left-moving electron-like, and right moving hole-like quasiparticles, respectively.

Below the gap, $\abs{E}<\Delta$, there are only two evanescent solutions, such that the most general form of the wave functions associated with the Hamiltonian \eqref{eq:sup:HS} reads
\begin{equation}
\psi(x) = \left[B_{\rm h} \viervector{-\alpha}{1}{0}{0} + B_{\rm e} \viervector{0}{0}{1}{-\alpha}\right]e^{\kappa x}.
\label{eq:sup:bs_L}
\end{equation}
Here $\alpha=(E-i\sqrt{\Delta^2-E^2})/\Delta$, which may be written as $\alpha = e^{-i \chi}$, where $\chi \in \reals$ is the phase shift acquired in an Andreev reflection process, and $\kappa = \sqrt{\Delta^2-E^2}/{v_{\rm{F}}}$ gives the decay length in the lead.

In the right (p-wave) lead, $x>L$, the Hamiltonian (2) reduces to 
\begin{equation}
{\cal H}_{\rm p} = v_{\text{F}} p \eta_z\tau_z -\Delta \eta_z\tau_xe^{-i\tau_z\phi}.
\label{eq:sup:HP}
\end{equation}
We notice that 
\begin{equation}
{\cal H}_{\rm p} = U^\dagger {\cal H}_{\rm s} U,\qquad {\rm where}\qquad U=\exp\left[i\tau_z\left(\frac\phi2+\frac\pi 4(1+\eta_z)\right)\right].
\label{eq:sup:Hp}
\end{equation}
This allows us to write the general form of the wave functions both in the continuum,
\begin{equation}
\psi(x) = \frac1{\sqrt{1+\alpha^2}}
\left[
C_{\rm e}^{\rm out} \viervector{1}{\alpha e^{-i\phi}}{0}{0}e^{i k x} 
+ C_{\rm h}^{\rm in} \viervector{\alpha e^{i\phi}}{1}{0}{0}e^{-i k x}+ C_{\rm e}^{\rm in} \viervector{0}{0}{1}{-\alpha e^{-i\phi}}e^{-i k x} +C_{\rm h}^{\rm out} \viervector{0}{0}{-\alpha e^{i\phi}}{1}e^{i k x} 
\right],
\label{eq:sup:wavefunction_R}
\end{equation}
and below the gap,
\begin{equation}
\psi(x) = \left[D_{\rm e} \viervector{1}{\alpha e^{-i\phi}}{0}{0} + D_{\rm h} \viervector{0}{0}{-\alpha e^{i\phi}}{1}\right]e^{-\kappa x}.
\label{eq:sup:bs_R}
\end{equation}
To determine the coefficients in the wave functions introduced above, we need to match them at the junction. For this, we derive the transfer matrix $M$ associated with the scalar potential $U(x)=U_0\theta[x(L-x)]$ in the normal part of the junction. When $U_0$ is large, the wave functions with energy $E$ in the normal part of the junction, $0<x<L$, are readily obtained as
\begin{equation}
\psi(x) = 
E_{\rm e}^{<} \viervector{1}{0}{-i}{0}e^{-\lambda x} 
+E_{\rm e}^> \viervector{1}{0}{i}{0}e^{\lambda x} 
+E_{\rm h}^< \viervector{0}{i}{0}{1}e^{-\lambda x} 
+E_{\rm h}^> \viervector{0}{-i}{0}{1}e^{\lambda x} 
,
\label{eq:sup:wavefunction_N}
\end{equation}
where $\lambda=U_0/v_{\rm F}$. Using the continuity conditions for the wave functions at $x=0$ and $x=L$, we can get rid of the coefficients $E_{\rm e}^<, E_{\rm e}^>, E_{\rm h}^<, E_{\rm h}^>$, and establish the relation 
\begin{equation}
\psi(L) = M \psi(0),
\label{eq:sup:matching}
\end{equation}
where $M=\cosh(\lambda L)+\sinh(\lambda L) \eta_y$. The coefficients in the transfer matrix can be related to the junction transparency, $T=1/\cosh^2(\lambda L)$. (For definiteness, we will assume $\lambda>0$ below.) At $T=0$, the two superconductors are decoupled. In that case, the boundary condition Eq.~\eqref{eq:sup:matching} reduces to 
\begin{eqnarray}
(1+\eta_y)\psi(0) &=& 0, \\
(1-\eta_y)\psi(L) &=& 0.
\end{eqnarray}
Below we use the matching condition \eqref{eq:sup:matching} to obtain the bound state and continuum wave functions. Furthermore, we consider the short-junction limit, $L \rightarrow 0$, while keeping the product $U_0L$ that determines the transparency constant.

\subsection{Bound state wave functions}
\label{sec:WFbound} 
In the transparent case, $T=1$, the matching equation provides two solutions in the R and L sectors, respectively. The solution in the R sector has energy 
\begin{equation}
E_{\rm R}=\Delta\sin\frac \phi 2 \sgn{\sin \phi};
\label{eq:sup:eR}
\end{equation}
its wave function is obtained with $D_{\rm h}=B_{\rm e}=0$ and $D_{\rm e}=ie^{i\phi/2} \sgn{\sin\phi}B_{\rm h}$. Using the normalization condition for the wave function, we can fix $B_{\rm h}=\sqrt{\Delta|\cos(\phi/2)|/(2v_{\rm F})} = \sqrt{\kappa_{\rm R}/2}$.    
The solution in the $L$ sector has energy 
\begin{equation}
E_{\rm L}=\Delta\cos\frac \phi 2;
\label{eq:sup:eL}
\end{equation}
its wave function is obtained with $D_{\rm e}=B_{\rm h}=0$ and $D_{\rm h}=-e^{-i\phi/2} B_{\rm e}$, where $B_{\rm e}=\sqrt{\Delta|\sin(\phi/2)|/(2v_{\rm F})}=\sqrt{\kappa_{\rm L}/2}$. 

The two states cross at $\phi=\pi/2$ and $\phi=3\pi/2$. The connection to the energy $E_+$ ($E_-$) given in the main text is made by taking for each intervall the state with the higher (lower) absolute value of the energy, i.e.,
\begin{eqnarray}
E_+(\phi) &=& \sgn{\sin\phi} \max \{ \abs{E_{\rm R}(\phi)}, \abs{E_{\rm L}(\phi)} \}, \\
E_-(\phi) &=& \sgn{\sin\phi} \min \{ \abs{E_{\rm R}(\phi)}, \abs{E_{\rm L}(\phi)} \}. 
\end{eqnarray}
At finite backscattering, these solutions hybridize and an avoided crossing appears near the phases $\phi=\frac{\pi}{2}$ and $\phi=\frac{3\pi}{2}$. In the opaque case, $T=0$, the higher-energy state merges with the continuum while the matching equation provides a unique bound state solution with energy $E_-=0$ that resides on the right side of the junction only. The coefficients are given as $B_{\rm h}=B_{\rm e}=0$ and $D_{\rm h}=e^{-i\phi}D_{\rm e}$ with $D_{\rm e}=\sqrt{\Delta/(2v_{\rm F})}=\sqrt{\kappa_-/2}$. 

At arbitrary transmission, we find two eigenstates with energies given by Eq.~(3) in the main text. Using the matching condition, Eq.~\eqref{eq:sup:matching}, and the normalization condition, $\int dx \absq{\Psi(x)} = 1$, we obtain the coefficients for the bound state with energy $E_{\nu}$:
\begin{eqnarray}
B_{\rm e}^{\nu} &=& \sqrt{T} \cos\left(\chi_{\nu} + \frac{\phi}{2}\right)  C^{\nu}, \\
B_{\rm h}^{\nu} &=& -i \sqrt{T(1-T)} \cos\frac{\phi}{2} C^{\nu}, \\
D_{\rm e}^{\nu} &=& \sqrt{1-T} e^{i\frac{\phi}{2}} \sin \chi_{\nu} C^{\nu}, \\
D_{\rm h}^{\nu} &=& \left[(1-T)\cos\frac{\phi}{2}-e^{-i \chi_{\nu}}\cos\left(\chi_{\nu}+\frac{\phi}{2}\right)\right] C^{\nu},
\end{eqnarray}
where
\begin{equation}
C^{\nu} = \sqrt{\frac{\kappa_{\nu}}{2 \cos 2\chi_{\nu} (T\cos^2\frac{\phi}{2}-\sin^2\chi_{\nu})}}.
\end{equation}
Note, that the expressions previously given for the special cases $T=0$ and $T=1$ differ by an irrelevant global phase factor.

\subsection{Continuum wave functions}
\label{sec:WFcontinuum} 

For a fixed energy in the continuum, $|E|>\Delta$, the relation between the four incoming and four outgoing wave functions encoded in Eqs.~\eqref{eq:sup:wavefunction_L} and \eqref{eq:sup:wavefunction_R} can be expressed through a scattering matrix $S(E)$ such that
\begin{equation}
\left(\begin{array}{c} A_{\rm e}^{\rm in} \\ C_{\rm e}^{\rm in} \\ A_{\rm h}^{\rm in} \\C_{\rm h}^{\rm in} \end{array} \right) = S^{-1}(E) \left(\begin{array}{c} A_{\rm e}^{\rm out} \\C_{\rm e}^{\rm out} \\ A_{\rm h}^{\rm out}\\ C_{\rm h}^{\rm out}  \end{array}\right) .
\label{eq:sup:defS-1}
\end{equation}
The scattering matrix is unitary, i.e., $S^{-1} = S^{\dagger}$.
At energies $|E|\gg\Delta$, the scattering matrix simplifies to $S=-i\sqrt{1-T}\tau_z+\sqrt{T}\eta_x$, in agreement with the transfer matrix introduced in Eq.~\eqref{eq:sup:matching}.

For a transparent junction, $T=1$, the scattering matrix is block diagonal as the R and L sectors decouple. It reads
\begin{equation}
S = \left(\begin{matrix}
0 & \frac{1-\alpha^2}{1-\alpha^2 e^{i\phi}} & \frac{\alpha (1-e^{i\phi})}{1-\alpha^2 e^{i\phi}} & 0 \\
\frac{1-\alpha^2}{1 + \alpha^2e^{-i\phi}} & 0 & 0 & -\frac{\alpha (1+e^{i\phi})}{1+\alpha^2e^{-i\phi}} \\
\frac{\alpha (1+e^{-i\phi})}{1+\alpha^2e^{-i\phi}} & 0 & 0 & \frac{1-\alpha^2}{1 + \alpha^2e^{-i\phi}} \\ 
0 & \frac{\alpha(e^{-i\phi}-1)}{1-\alpha^2 e^{i\phi}} & \frac{1-\alpha^2}{1-\alpha^2 e^{i\phi}} & 0
\end{matrix}\right).
\label{eq:sup:sT1}
\end{equation}
For the opaque junction, $T=0$, scattering states are confined within each lead, and the scattering matrix reads
\begin{equation}
S = \left( \begin{matrix}
-i & 0 & 0 & 0 \\
0 & i \frac{\alpha^2-1}{1+\alpha^2} & 0 & -\frac{2\alpha}{1+\alpha^2}e^{i\phi} \\
0 & 0 & i & 0 \\
0 & \frac{2\alpha}{1+\alpha^2}e^{-i\phi} & 0 & -i \frac{\alpha^2 -1}{1+\alpha^2}
\end{matrix} \right) = 
\left(\begin{array}{cccc} 
-i&0&0&0\\
0&i\sqrt{1-\frac{\Delta^2}{E^2}} &0&-\frac\Delta  E e^{i\phi}\\
0&0&i&0\\
0&\frac\Delta  E e^{-i\phi}&0&-i\sqrt{1-\frac{\Delta^2}{E^2}}
\end{array}\right),
\label{eq:sup:S0}
\end{equation}
where we used Eqs.~\eqref{eq:sup:alpha} in the last step.

In the general case, using the matching condition in Eq.~\eqref{eq:sup:matching} and the continuum wave functions in Eqs.~\eqref{eq:sup:wavefunction_L} and \eqref{eq:sup:wavefunction_R}, the inverse scattering matrix can be written as $S^{-1} = B^{-1} A $ with
\begin{equation}
A = \left( \begin{matrix}
0 & -1 & -\alpha \sqrt{T} & -i \sqrt{1-T} \alpha e^{i\phi} \\
0 & -\alpha e^{-i\phi} & \sqrt{T} & i \sqrt{1-T} \\
\sqrt{T} & -i \sqrt{1-T} & 0 & \alpha e^{i\phi} \\
-\alpha \sqrt{T} & -i \sqrt{1-T} \alpha e^{-i\phi} & 0 & -1 
\end{matrix}\right),
\end{equation}
\begin{equation}
B = \left(\begin{matrix}
- \sqrt{T} & -i \sqrt{1-T} & 0 & \alpha e^{i\phi} \\
\alpha \sqrt{T} & i \alpha \sqrt{1-T} e^{-i\phi} & 0 & 1 \\
0 & 1 & \alpha \sqrt{T} & i \sqrt{1-T} \alpha e^{i\phi} \\
0 & -\alpha e^{-i\phi} & - \sqrt{T} & i \sqrt{1-T}
\end{matrix}\right).
\end{equation}
In the following, we will use the outgoing continuum states. Their wave function is obtained by setting one of the outgoing coefficients $A_{\rm e}^{\rm out}, C_{\rm e}^{\rm out}, A_{\rm h}^{\rm out}, C_{\rm h}^{\rm out}$ to unity and computing the incoming coefficients via the scattering matrix, Eq.~\eqref{eq:sup:defS-1}.

\section{transition rates}
\label{sec:rates}

In this section we calculate the ionization and refill rates induced by a weak ac Zeeman field using second order perturbation theory. For this, we first derive the Hamiltonian due to the Zeeman field, Eq.~(9) in the main text, in the unperturbed basis of the wave functions introduced in Sec.~\ref{sec:wavefunctions}. Introducing the Bogoliubov transformation,
\begin{equation}
\Psi(x)=\sum_\lambda\psi_\lambda(x)\gamma_\lambda,
\end{equation}
inserting it into Eq. (8) in the main text, and symmetrizing the resulting expression, we obtain
\begin{equation}
\label{eq:hz1}
H_{\rm Z}=\frac{\mu_{\rm B}B}2\sum_{\lambda\lambda'}V_{\lambda,\lambda'}\gamma_\lambda\gamma_{\lambda'} + {\rm h.c.},
\end{equation}
where $V_{\lambda,\lambda'}$ is given below Eq. (9) in the main text. Note that $V_{\lambda,\lambda'}=-V_{\lambda',\lambda}$. This allows writing Eq. \eqref{eq:hz1} as
\begin{eqnarray}
H_{\text{Z}} &=& \mu_{\text{B}} Be^{-i \Omega t} 
\left(V_{+,-} \gamma_+ \gamma_- + \sum_{E;\mu,\nu} V_{\nu,E\mu} \gamma_{\nu} \gamma_{E\mu}  
+ \frac 12 \sum_{E,E';\mu,\mu'} V_{E\mu,E'\mu'}\gamma_{E\mu}\gamma_{E'\mu'}\right) + \mathrm{h.c.}
\label{eq:HBqp}
\end{eqnarray}
Using the definition of ionization and refill rates in the main text, we can calculate them by applying Fermi's golden rule to the Hamiltonian \eqref{eq:HBqp},
\begin{eqnarray}
\Gamma^{{\rm I/R}}_{\nu} (\Omega)= 2\pi (\mu_{\rm{B}} B)^2 \sum_{E,\mu}  \absq{V_{\nu,E\mu}} 
\delta(\Omega + E_{\nu} + E)\, \theta ( \mp   E  E_\nu).
\end{eqnarray} 
Here, the upper sign is for an ionization process, whereas the lower sign is for a refill process. The Heaviside function appears due to the Fermi-Dirac distributions at zero temperature, ensuring that in a refill process the bound state and the continuum state are empty, and in an ionization process the bound state is occupied whereas the continuum state is empty. Noting that ${\rm sign}( E_\nu)={\rm sign}( \sin\phi)$, we obtain
\begin{equation}
\Gamma^{{\rm I/R}}_{\nu} (\Omega)= 2\pi (\mu_{\rm{B}} B)^2 \sum_{E'>0,\mu}  \absq{V_{\nu;\mp E'{\rm sign}( \sin\phi ),\mu}} 
\delta[\Omega + (|E_{\nu}|\mp E'){\rm sign}( \sin\phi)].
\label{eq:sup:rates}
\end{equation} 
Using the density of states in the leads $\rho(E)$, we can replace the sum by an integral and obtain Eq.~(17) in the main text.

To obtain the matrix elements $V_{\nu, E\mu}$, we use the general expressions for the bound state wave functions and the continuum wave functions, defined above in Eqs.~\eqref{eq:sup:bs_L}, \eqref{eq:sup:bs_R} and \eqref{eq:sup:wavefunction_L}, \eqref{eq:sup:wavefunction_R}, respectively. After integration over the real space coordinate and reorganization, we obtain
\begin{equation}
V_{\nu,E\mu} = \frac{1}{\sqrt{1+\alpha^2}} \left( \frac{F_1^{\nu}}{\kappa_{\nu} + ik} +\frac{F_2^{\nu}}{\kappa_{\nu} - ik} \right),
\label{eq:sup:V}
\end{equation}
where
\begin{eqnarray}
F_1^{\nu} &=& (B_{\rm e} A_{\rm e}^{\rm in} -B_{\rm h} A_{\rm h}^{\rm in})(\alpha - e^{-i\chi_{\nu}})+ (D_{\rm h} C_{\rm h}^{\rm in} e^{i\phi} +D_{\rm e} C_{\rm e}^{\rm in} e^{-i\phi})(\alpha + e^{-i\chi_{\nu}}), \\
F_2^{\nu} &=& (B_{\rm e}A_{\rm h}^{\rm out}-B_{\rm h} A_{\rm e}^{\rm out}) (\alpha e^{-i\chi_{\nu}} -1) + (D_{\rm h} C_{\rm e}^{\rm out}-D_{\rm e} C_{\rm h}^{\rm out}) (1+\alpha e^{-i\chi_{\nu}}).
\end{eqnarray}
Using the outgoing wave functions as defined above, Eq.~\eqref{eq:sup:V} can be evaluated numerically to obtain the rates for arbitrary transmission. 
In the following, we consider the two special cases of a transparent and an opaque junction, for which we give the analytical expressions.

\subsection{Transparent junction}
Here, we want to calculate the rates given by Eq.~(17) in the main text for $T=1$.
We have seen, that there are two bound states, labelled by R and L, with energies given in Eqs.~\eqref{eq:sup:eR} and \eqref{eq:sup:eL}. For each of them, we need to evaluate \eqref{eq:sup:V}. Since the calculation is very similar in both cases, we will only show the explicit calculation for $\nu = {\rm L}$. Using the coefficients for $T=1$, given below Eq.~\eqref{eq:sup:eL}, and $\chi_{\rm L} = \phi/2$, we find
\begin{align}
V_{{\rm L},E\mu} &= \sqrt{\frac{\kappa_{\rm L}}{2(1+\alpha^2)}} \left[ \frac{A_{\rm e}^{\rm in}\left(\alpha - e^{-i\frac{\phi}{2}}\right)-C_{\rm h}^{\rm in}\left(\alpha e^{i\frac{\phi}{2}} + 1\right)}{\kappa_{\rm L} + ik}   + \frac{A_{\rm h}^{\rm out}\left(\alpha  e^{-i\frac{\phi}{2}} -1\right)-e^{-i\frac{\phi}{2}}C_{\rm e}^{\rm out} \left(1+ \alpha e^{-i\frac{\phi}{2}}\right)}{\kappa_{\rm L} - ik} \right].
\end{align}
For the transparent junction, there are only two outgoing states, $C_{\rm e}^{\rm out} = 1$ or $A_{\rm h}^{\rm out}=1 $. Then,
\begin{align}
\begin{split}
\sum_{\mu} \absq{V_{{\rm L},E \mu}} 
 &= \frac{\kappa_{\rm L}}{2(1+\alpha^2)} \left\{\frac{2(1+\alpha^2)}{\kappa_{\rm L}^2 + k^2} + \frac{\alpha^2 - 2\alpha \epsilon_{\rm L} +1}{\kappa_{\rm L}^2 + k^2}\left(\absq{S_{21}} + \absq{S_{31}} \right) + \frac{\alpha^2 + 2\alpha \epsilon_{\rm L} +1}{\kappa_{\rm L}^2 + k^2} \left( \absq{S_{24}} + \absq{S_{34}} \right)  \right. \\
-& \left.  \frac{2}{\kappa_{\rm L}^2 + k^2} \Re \left[ (\alpha - e^{-i\frac{\phi}{2}} ) (1 +\alpha e^{-i\frac{\phi}{2}}) (S_{21}^* S_{24} + S_{31}^* S_{34}) \right] + 2 \Re \left[ \frac{1}{(\kappa_{\rm L} + ik)^2} H_1  \right]  + 2 \Re \left[ \frac{1}{(\kappa_{\rm L} - ik)^2} H_2 \right] \right\}.
\end{split}
\end{align}
where
\begin{eqnarray}
H_1 &=& (\alpha -e^{-i\frac{\phi}{2}}) (-S_{21}^{*} (\alpha e^{-i\phi} +  e^{i\frac{\phi}{2}}) + S_{31}^{*} (-1 +\alpha  e^{i\frac{\phi}{2}} )),\\
H_2 &=& -(1 +\alpha e^{-i\frac{\phi}{2}}) (-S_{24} (\alpha e^{-i\phi} + e^{-i\frac{\phi}{2}}) + S_{34} (-1 +\alpha  e^{-i\frac{\phi}{2}} )).
\end{eqnarray}
Using the unitarity of the scattering matrix, $\sum_k S_{ik} S_{jk}^* = \delta_{ij}$, we find
\begin{equation}
\sum_{\mu} \absq{V_{{\rm L},E\mu}} = \frac{2\kappa_{\rm L}}{\kappa_{\rm L}^2 + k^2} + \frac{\kappa_{\rm L}}{1+\alpha^2} \left( \frac{\kappa_{\rm L}^2 - k^2}{(\kappa_{\rm L}^2+k^2)^2} \Re [H_1 + H_2] + \frac{2\kappa_{\rm L} k}{(\kappa_{\rm L}^2 + k^2)^2 } \Im [H_1 - H_2]   \right).
\end{equation}
which after lengthy, but straightforward, algebraic manipulation yields
\begin{align}
\begin{split}
\sum_{\mu} \absq{V_{{\rm L},E\mu}} &= \frac{2\kappa_{ \rm L}}{\kappa_{\rm L}^2 +k^2} \left[ 1 +\frac{\kappa_{\rm L}^2 -k^2}{\kappa_{\rm L}^2 + k^2} \frac{1-4\alpha^2+\alpha^4-2\alpha^2 \cos\phi}{1+\alpha^4 +2\alpha^2 \cos\phi} + \frac{\kappa_{\rm L} k}{\kappa_{\rm L}^2 +k^2} \frac{8 \alpha^2 (\alpha^2 -1)\sin\phi}{(1+\alpha^2)(1+\alpha^4 + 2\alpha^2 \cos\phi)} \right]. \\
&= \frac{v_{\rm F}}{\Delta} \frac{4\abs{\sin\frac{\phi}{2}}(\epsilon^2-1)}{(\epsilon^2-\cos^2\frac{\phi}{2})^2(\epsilon^2-\sin^2\frac{\phi}{2})}\left[ 1- \frac{\sin\phi \abs{\sin\frac{\phi}{2}}}{\epsilon}\right].
\end{split}
\end{align}
where $\epsilon= E / \Delta$.
Repeating the calculation for $\nu = {\rm R}$ and using the expressions for the bound state energies, Eqs. \eqref{eq:sup:eR} and \eqref{eq:sup:eL}, we finally obtain
\begin{equation}
\sum_{\eta} \absq{V_{{\nu},\eta}} = \frac{v_{\rm F}}{\Delta} \frac{4\abs{\epsilon_{\bar{\nu}}}(\epsilon^2-1)}{(\epsilon^2-\epsilon_{\nu}^2)^2(\epsilon^2-\epsilon_{\bar{\nu}}^2)}\left[ 1- \frac{\sin\phi \abs{\epsilon_{\bar{\nu}}}}{\epsilon}\right],
\label{eq:sup:Vq} 
\end{equation}
where $\epsilon_{\nu}=E_{\nu}/\Delta$ and $\bar{R}=L$ and $\bar{L}=R$. Substituting \eqref{eq:sup:Vq} into \eqref{eq:sup:rates} and using the energy conservation condition, $|\epsilon|= |\tilde{\Omega}| + |\epsilon_{\nu}|$ for an ionization process and $|\epsilon|= |\tilde{\Omega}| - |\epsilon_{\nu}|$ for a refill process, we find the rates
\begin{equation}
\Gamma_{\nu}^{\rm I/R}= \frac{(\mu_{\rm B} B)^2}{\Delta} \frac{4 \abs{\epsilon_{\bar{\nu}}}\sqrt{(|\tilde{\Omega}|\pm \abs{\epsilon_{\nu}})^2 -1}[|\tilde{\Omega}|\pm \abs{\epsilon_{\nu}}+\sgn{\Omega}\abs{\epsilon_{\bar{\nu}}}\sin\phi]}{|\tilde{\Omega}|^2 (|\tilde{\Omega}|\pm 2\abs{\epsilon_{\nu}})^2[(|\tilde{\Omega}| \pm \abs{\epsilon_{\nu}})^2-\absq{\epsilon_{\bar{\nu}}}]},
\end{equation}
where we defined $\tilde{\Omega} = \Omega/\Delta$. 

Near the threshold frequency, $|\tilde\Omega_{\nu, \rm c}^{\rm I/R}|= 1 \mp \abs{\epsilon_{\nu}}$, 
the rates $\Gamma_{\nu}^{\rm I/R}$ grow as 
$\sqrt{\delta\tilde \Omega}$, where $\delta\tilde \Omega=|\tilde \Omega|-|\tilde\Omega_{ \nu,\rm c}^{\rm I/R}|$.  At large frequencies they decrease as $1/\tilde\Omega^4$. 
To describe the intermediate frequency regime, we concentrate on the regime $\phi,\delta\tilde \Omega\ll1$ (a similar situation occurs for phases $\phi$ close to $\pi$), where
\begin{equation}
\Gamma_-^{\rm I/R} = \frac{(\mu_{\rm B} B)^2}{\Delta}\frac{2 \sqrt{2}\sqrt{\delta \tilde{\Omega}}}{\delta \tilde{\Omega} + \frac{ \phi ^2}{8}}\qquad {\rm and}\qquad \Gamma_+^{\rm I/R} = \frac{(\mu_{\rm B} B)^2}{\Delta}\frac{\sqrt{\delta \tilde{\Omega}} \phi}{\sqrt{2}(\delta \tilde{\Omega} + \frac{\phi ^2}{8})^2}.
\end{equation}
Thus we find that the former reaches its maximum, $\Gamma_{-,\rm max}/[(\mu_{ \rm B} B)^2/\Delta]=4/\phi$, at $|\delta \tilde \Omega_{-,\rm max}|=\phi^2/8$, while the later reaches its maximum, $\Gamma_{+,\rm max}/[(\mu_{ \rm B} B)^2/\Delta]=6\sqrt{3}/\phi^2$, at $|\delta\tilde \Omega_{+,\rm max}|=\phi^2/24$.

At larger $\phi$, the maximum is less pronounced and further away from the threshold than for small $\phi$. Note, that for $\phi=\pi / 2$, $\Gamma^{\rm I/R}_+ = \Gamma^{\rm I/R}_-$, since the bound states are degenerate.

\subsection{Opaque junction}
Here, we evaluate \eqref{eq:sup:V} for the opaque junction, $T=0$. Using the coefficients derived in Section \ref{sec:WFbound}, we obtain
\begin{equation}
V_{-,E\mu} = e^{-i\phi}\sqrt{\frac{\kappa_-}{2(1+\alpha^2)}} \left[-i \frac{C_{\rm h}^{\rm in} e^{i\phi}+C_{\rm e}^{\rm in}}{\kappa_- + ik}(1+i\alpha) +\frac{C_{\rm e}^{\rm out} - C_{\rm h}^{\rm out} e^{i\phi}}{\kappa_- - ik} (1-i \alpha) \right].
\end{equation} 
As in the transparent case, there are only two outgoing states, $C_{\rm e}^{\rm out} = 1$ or $C_{\rm h}^{\rm out}=1 $. 
After some algebra and using the unitarity of the scattering matrix, we find
\begin{equation}
\sum_{\mu}\absq{V_{-,E\mu}} = \frac{\kappa_-}{1+\alpha^2}  \left\{2 \frac{1+\alpha^2}{\kappa_-^2 + k^2} -\Im \left[ \frac{(1-i\alpha)^2}{(\kappa_- -ik)^2} (S_{22}-S_{24}+e^{-i\phi}S_{44}-e^{i\phi}S_{42}) \right] \right\}.
\end{equation}
Using the scattering matrix \eqref{eq:sup:S0}, we finally obtain
\begin{equation}
\sum_{\mu}\absq{V_{-,E\mu}} = \frac{v_{\rm F}}{\Delta} \frac{16(\epsilon^2-1)}{\epsilon^6}.
\end{equation}
Since the bound state energy is zero, the rate is identic for refill and ionization processes and reads
\begin{equation}
\Gamma = \frac{(\mu_{ \rm B} B)^2}{\Delta} \frac{16 \sqrt{|\tilde{\Omega}|^2-1}}{|\tilde{\Omega}|^5}.
\end{equation}
Near the threshold frequency, $|\tilde{\Omega}_{\rm c}|=1$, the rate grows as $\sqrt{\delta \tilde{\Omega}}$. It reaches its maximum, $\Gamma_{\rm max}/[(\mu_{ \rm B} B)^2/\Delta]=2^8/(25\sqrt{5})$, at $|\tilde \Omega_{\rm max}|=\sqrt{5}/2$ and decreases as $1/\tilde\Omega^4$ at large frequencies.
Note, that the rate does not depend on the superconducting phase difference, since the bound state energy is independent of the phase.

\section{stationary occupations}
\label{sec:stationary}
In this section we give the analytical expressions for the stationary occupations $\rho^{\rm st}_{ii}$ of the state $|i\rangle$, that are used for the calculation of the supercurrent. These stationary occupations are obtained from the master equation, Eq.~(11) in the main manuscript, by setting $\dot{\rho}=0$.
As the current in the main text, they are most conveniently expressed in the form 
\begin{equation}
\rho^{\rm st}_{ii}=\rho^\infty_{ii}+\frac{\Gamma^2}{\Gamma^2+(2\delta\omega)^2}(\rho^0_{ii}-\rho^\infty_{ii}).
\end{equation} 
The occupations far from resonance ($\rho^{\infty}_{ii}$) are given as 
\begin{equation}
\rho_{00}^{\infty} = \frac{\Gamma_+^{\rm{I}} \Gamma_-^{\rm{I}}}{\Gamma_+ \Gamma_-}, \quad
\rho_{--}^{\infty} = \frac{\Gamma_+^{\rm{I}} \Gamma_-^{\rm{R}}}{\Gamma_+ \Gamma_-}, \quad
\rho_{++}^{\infty} = \frac{\Gamma_-^{\rm{I}} \Gamma_+^{\rm{R}}}{\Gamma_+ \Gamma_-}, \quad
\rho_{22}^{\infty} = \frac{\Gamma_+^{\rm{R}} \Gamma_-^{\rm{R}}}{\Gamma_+ \Gamma_-},
\end{equation}
whereas the occupations at the resonance ($\rho^{0}_{ii}$) take the form
\begin{eqnarray}
\rho_{00}^0 &=& \rho_{00}^{\infty} \left\{ 1+ \frac{\absq{\omega_1}}{\Gamma^2} \left[ \left(1+\frac{\Gamma_-^{\rm{R}}}{\Gamma_+^{\rm{I}}}\right)\left(1+\frac{\Gamma_+^{\rm{R}}}{\Gamma_-^{\rm{I}}}\right)-\frac{\Gamma_{\Sigma}^2}{\Gamma_+ \Gamma_-} \right] \right\} ,
\label{eq:sup:rho_1}\\
\rho_{--}^0 &=& \rho_{--}^{\infty} \left\{ 1+ \frac{\absq{\omega_1}}{\Gamma^2} \left[ \frac{(\Gamma_+^{\rm{I}} + \Gamma_-^{\rm{R}})^2}{\Gamma_+^{\rm{I}} \Gamma_-^{\rm{R}}}-\frac{\Gamma_{\Sigma}^2}{\Gamma_+ \Gamma_-} \right] \right\},\\
\rho_{++}^0 &=& \rho_{++}^{\infty} \left\{ 1 + \frac{\absq{\omega_1}}{\Gamma^2} \left[ \frac{(\Gamma_-^{\rm{I}} + \Gamma_+^{\rm{R}})^2}{\Gamma_-^{\rm{I}}\Gamma_+^{\rm{R}}} - \frac{\Gamma_{\Sigma}^2}{\Gamma_+ \Gamma_-}\right] \right\},\\
\rho_{22}^0 &=& \rho_{22}^{\infty} \left\{ 1+\frac{\absq{\omega_1}}{\Gamma^2} \left[ \left(1+\frac{\Gamma_+^{\rm{I}}}{\Gamma_-^{\rm{R}}}\right)\left(1+\frac{\Gamma_-^{\rm{I}}}{\Gamma_+^{\rm{R}}}\right)-\frac{\Gamma_{\Sigma}^2}{\Gamma_+ \Gamma_-} \right] \right\},
\label{eq:sup:rho_4} 
\end{eqnarray}
The expressions for the current given in Eqs.~(12) and (13) in the main manuscript are obtained by inserting the stationary occupations, Eqs.~\eqref{eq:sup:rho_1} - \eqref{eq:sup:rho_4}, into Eq.~(10) in the main manuscript.